%% file: main.tex
\documentclass[sigconf]{acmart}

\usepackage{microtype}
\usepackage{graphicx}
\usepackage{subfigure}
\usepackage{booktabs} % for professional tables
\usepackage{subcaption}

\usepackage{hyperref}
\usepackage{multirow}
\usepackage{xspace}
\usepackage{afterpage}
\newcommand{\name}{KnowGraph~}
\newcommand{\namenospace}{KnowGraph}

\usepackage{amsmath}
\usepackage{mathtools}
\usepackage{amsthm}
\usepackage{xcolor}
\usepackage[table]{colortbl}

\usepackage[capitalize,noabbrev]{cleveref}

\theoremstyle{plain}

\theoremstyle{definition}

\theoremstyle{remark}

\usepackage[textsize=tiny]{todonotes}

\usepackage{color}

\newcommand{\method}{KnowGraph}

\newcommand{\pT}{{\mathcal{T}}}

\AtBeginDocument{%
  }

\copyrightyear{2024}
\acmYear{2024}
\setcopyright{rightsretained}
\acmConference[CCS '24]{Proceedings of the 2024 ACM SIGSAC Conference on Computer and Communications Security}{October 14--18, 2024}{Salt Lake City, UT, USA}
\acmBooktitle{Proceedings of the 2024 ACM SIGSAC Conference on Computer and Communications Security (CCS '24), October 14--18, 2024, Salt Lake City, UT, USA}
\acmDOI{10.1145/3658644.3690354}
\acmISBN{979-8-4007-0636-3/24/10}

\acmSubmissionID{ccsfpb2556}

\begin{document}

%%
%% The "title" command has an optional parameter,
%% allowing the author to define a "short title" to be used in page headers.
\title{KnowGraph: Knowledge-Enabled Anomaly Detection via Logical Reasoning on Graph Data}

%%
%% The "author" command and its associated commands are used to define
%% the authors and their affiliations.
%% Of note is the shared affiliation of the first two authors, and the
%% "authornote" and "authornotemark" commands
%% used to denote shared contribution to the research.
%\author{}
\author{Andy Zhou}
\affiliation{%
  \institution{UIUC}
  \country{}
  % \institution{Lapis Labs}
  \city{Champaign, IL}
  \country{USA}
}
\email{andyz3@illinois.edu}

\author{Xiaojun Xu}
\affiliation{%
  \institution{Bytedance Research}
  \country{}
  \city{San Jose, CA}
  \country{USA}
}
\email{xiaojun.xu@bytedance.com}

\author{Ramesh Raghunathan}
\affiliation{%
  \institution{eBay}
  \country{}
  \city{Austin, TX}
  \country{USA}
  }
\email{raraghunathan@ebay.com}

\author{Alok Lal}
\affiliation{%
  \institution{eBay}
  \country{}
  \city{San Jose, CA}
  \country{USA}
  }
\email{allal@ebay.com}

\author{Xinze Guan}
\affiliation{%
  \institution{eBay}
  \country{}
  \city{San Jose, CA}
  \country{USA}
}
\email{xiguan@ebay.com}

\author{Bin Yu}
\affiliation{%
  \institution{UC Berkeley}
  \country{}
  \city{Berkeley, CA}
  \country{USA}
  }
\email{binyu@berkeley.edu}

\author{Bo Li}
\affiliation{%
  \institution{UIUC}
  \country{}
  \city{Champaign, IL}
  \country{USA}
  }
\email{lbo@illinois.edu}

%%
%% By default, the full list of authors will be used in the page
%% headers. Often, this list is too long, and will overlap
%% other information printed in the page headers. This command allows
%% the author to define a more concise list
%% of authors' names for this purpose.
\renewcommand{\shortauthors}{Andy Zhou et al.}

%%
%% The abstract is a short summary of the work to be presented in the
%% article.
\begin{abstract}
Graph-based anomaly detection is pivotal in diverse security applications, such as fraud detection in transaction networks and intrusion detection for network traffic. Standard approaches, including Graph Neural Networks (GNNs), often struggle to generalize across shifting data distributions. For instance, we observe that a real-world eBay transaction dataset revealed an over 50\% decline in fraud detection accuracy when adding data from only a single new day to the graph due to data distribution shifts. This highlights a critical vulnerability in purely data-driven approaches.
Meanwhile, real-world domain knowledge, such as ``simultaneous transactions in two locations are suspicious,'' is more stable and a common existing component of real-world detection strategies. To explicitly integrate such knowledge into data-driven models such as GCNs, we propose \namenospace, which integrates domain knowledge with data-driven learning for enhanced graph-based anomaly detection. \name comprises two principal components: (1) a statistical learning component that utilizes a main model for the overarching detection task, augmented by multiple specialized knowledge models that predict domain-specific semantic entities; (2) a reasoning component that employs probabilistic graphical models to execute logical inferences based on model outputs, encoding domain knowledge through weighted first-order logic formulas.
In addition, \name has leveraged the Predictability-Computability-Stability (PCS) framework for veridical data science to estimate and mitigate prediction uncertainties.
Empirically, \name has been rigorously evaluated on two significant real-world scenarios: collusion detection in the online marketplace eBay and intrusion detection within enterprise networks. 
Extensive experiments on these large-scale real-world datasets show that \name consistently outperforms state-of-the-art baselines in both transductive and inductive settings, achieving substantial gains in average precision when generalizing to completely unseen test graphs. 
% For instance, \name significantly improves baseline model performance, with a 0.151 inductive AUC gain on eBay collusion detection and 0.0139 inductive AUC gain on intrusion detection on LANL. 
Further ablation studies demonstrate the effectiveness of the proposed reasoning component in improving detection performance, especially under extreme class imbalance. These results highlight the potential of integrating domain knowledge into data-driven models for high-stakes, graph-based security applications.
\end{abstract}
%%
%% The code below is generated by the tool at http://dl.acm.org/ccs.cfm.
%% Please copy and paste the code instead of the example below.
%%

%%
%% Keywords. The author(s) should pick words that accurately describe
%% the work being presented. Separate the keywords with commas.
%% A "teaser" image appears between the author and affiliation
%% information and the body of the document, and typically spans the
%% page.

% \received{?????}
% \received[revised]{?????}
% \received[accepted]{?????}

%%
%% This command processes the author and affiliation and title
%% information and builds the first part of the formatted document.
\maketitle

\section{Introduction}

Graphs are ubiquitous data structures with applications in various domains, such as social networks, biological networks, and communication networks. In recent years, graph neural networks (GNNs) \cite{Zhou2018GraphNN, Wu2019ACS} and other graph representation learning techniques, such as node2vec \cite{Grover2016node2vecSF}, have emerged as powerful techniques for learning on graph-structured data. These methods have achieved remarkable success in applications that involve graphs such as recommendation \cite{Fan2019GraphNN}, drug discovery \cite{Zhang2022GraphNN}, NLP \cite{Park2022EvoKGJM}, traffic forecasting \cite{Jiang2021GraphNN}, and social network analysis \cite{Newman2002RandomGM}. In addition, graph data-based attacks have led to various real-world scenarios, such as fraudulent transactions in financial networks \cite{Cao2017HitFraudAB, Rao2020xFraudEF, Weber2019AntiMoneyLI,wu2022linkteller}, fake reviews or accounts in social networks \cite{Hooi2015BIRDNESTBI, Hinkelmann2022CombiningML, Breuer2020FriendOF}, and network intrusions in computer networks \cite{Liu2019Log2vecAH,freitas2020d2m,king2023euler}. Due to the large economic cost of these anomalies \cite{AlFalahi2019ConceptualBO, Mt2019THEEO}, it is crucial to develop automated and robust methods to detect them. 

\begin{figure}[t]
    \centering
    \hspace{-3mm} 
    \includegraphics[width=0.49\textwidth]{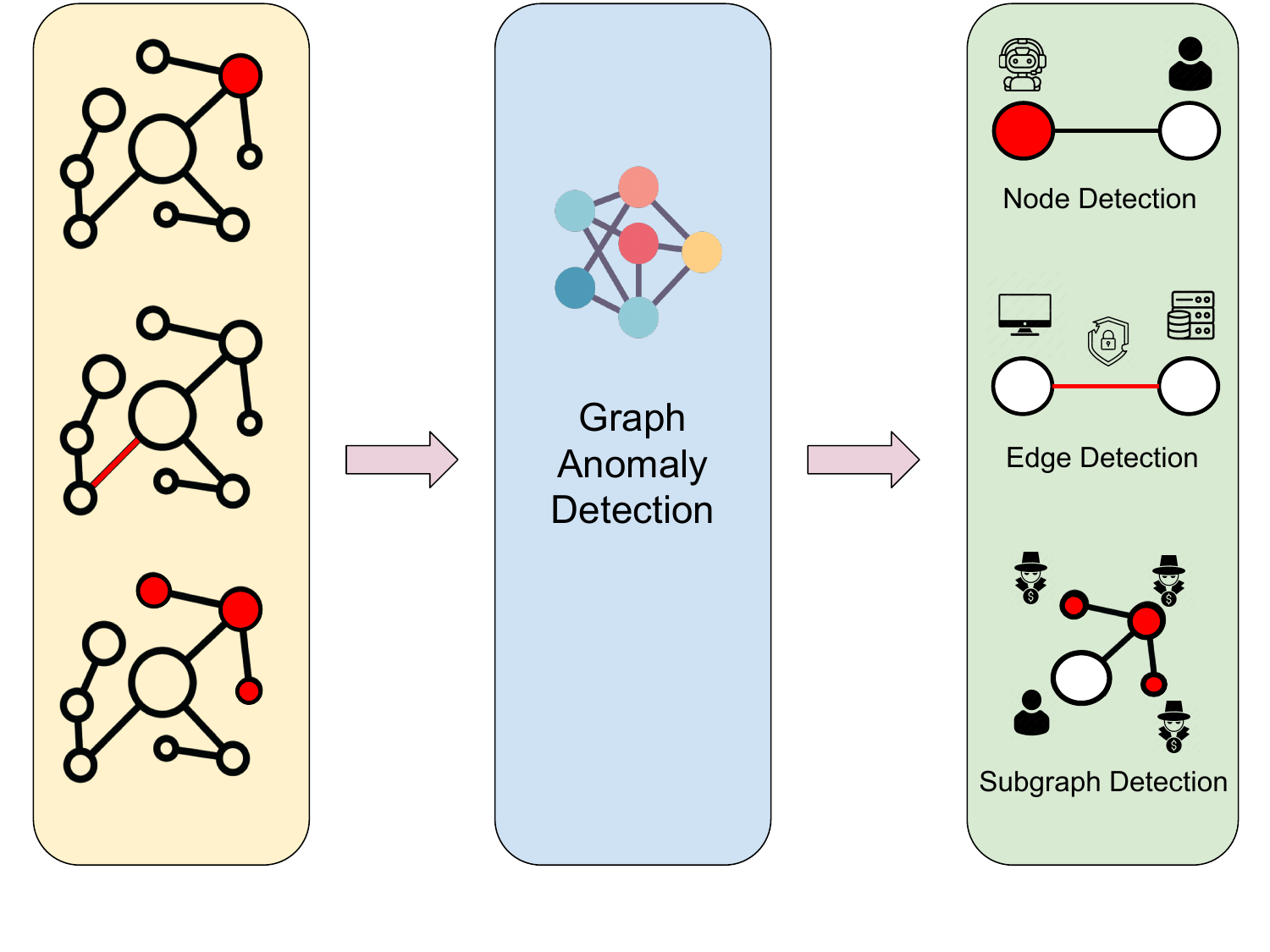}
    \vspace{-10mm}
    \caption{Examples of anomaly detection on graph-structure data with GNN models. Data-driven learning approaches have been successful on node-level, edge-level, and subgraph-level tasks but tend to consider different levels of the graph separately, focusing on a single level.}
    \label{fig:teaser}
    \vspace{-0.2in}
\end{figure}

Anomaly detection on graphs aims to identify these abnormal entities, connections, or substructures that deviate significantly from the expected patterns, summarized in Fig.~\ref{fig:teaser}. However, real-world graphs are often larger and more complex, holding properties such as extreme label imbalance and heterogeneity. For example, a single graph might have both locally homophilous and locally heterophilous regions \cite{zhu2020homophily}, making generalization, especially to unseen graphs in inductive settings, challenging. While recent techniques leverage the structural nature of network or marketplace data through GNNs \cite{Bowman2020DetectingLM, king2023euler, Weber2019AntiMoneyLI, Liu2018HeterogeneousGN, Ma2018GraphRADA}, several challenges remain when applied to large-scale real-world graphs, hindering their effectiveness in these critical domains.
For instance, when a GCN model is deployed on a real-world eBay marketplace transaction dataset to detect fraudulent activities, its accuracy drops by 50\% when only a single day's worth of new transactions is added to the graph, highlighting the challenge of maintaining performance on constantly evolving graphs.

% \bo{mention ebay example and emphasize GCN based approach hardly generalize in such real-world dataset}

This issue arises largely because current machine learning systems for anomaly detection are often purely data-driven, relying on the availability of large amounts of labeled data to learn patterns and anomalies \cite{Ma2021ACS}.  However, this approach faces several challenges. One major challenge is information heterogeneity \cite{Wang2019HeterogeneousGA}, as these techniques must leverage diverse sources of information, such as zip codes, item details, account data, authentication types, IP addresses, and event timestamps. Another challenge is label imbalance, as anomalous events or transactions are often extremely rare compared to benign ones, making it difficult to train ML systems that can reliably distinguish between them while maintaining a low false positive rate \cite{Ma2023ClassImbalancedLO}. Collecting labels for every graph component is also costly and challenging, especially for large-scale graphs with millions of nodes. These factors make inductive generalization especially challenging, which is crucial as real-world graphs constantly evolve with millions of new events daily.

In contrast, traditional intrusion detection approaches often rely on human-defined rules and heuristics \cite{Cohen1995FastER,Brause1999NeuralDM,Bohara2017AnUM,Rosset1999DiscoveryOF}. While these rules may not be as flexible as data-driven methods, they can help address some of the challenges faced by current ML systems. For instance, domain experts can provide insights into the most informative features for fraud detection or lateral movement, reducing the reliance on costly labeled data. They can also define logical rules that character the relationships between different entities in the graph, enhancing the interpretability and robustness of the model. Furthermore, incorporating domain knowledge can guide the model to focus on the most relevant subgraphs or patterns, improving its ability to generalize to unseen graphs. However, standard GNN architectures, such as GCN \cite{kipf2017semisupervised}, rely on global graph properties and cannot incorporate valuable human knowledge. Intuitively, integrating domain-specific knowledge into these architectures can significantly improve their generalization and interpretability.

In this paper, we propose a framework, \namenospace, that integrates expert knowledge into graph neural networks to bridge the gap between purely data-driven and rule-based approaches. Our framework consists of two key components: (1) a \textit{learning} component that utilizes a \textit{main model} for the overarching detection task, augmented by multiple specialized \textit{knowledge models} that are trained with diverse objectives and can predict domain-specific semantic entities; (2) a \textit{reasoning component} that utilizes probabilistic graphical models to perform logical inferences based on the outputs of each knowledge model. For statistical learning, we develop a set of complex models that can predict various aspects of graph data, such as node attributes, edge properties, and subgraph patterns. This expands the representational capacity of the main model, which is otherwise limited to a single overall facet. For reasoning, domain knowledge is encoded through first-order logic formulas with learned weights that organize each model's output, allowing the framework to learn how to leverage ground-truth information and constraints during inference. Due to the high-stakes nature of our experimental setting, we follow the Predictability, Computability, and Stability (PCS) framework \cite{Yu2020VeridicalDS, Yu2023WhatIsUncertainty, Yu2013Stability, Yu2024Book} for our evaluation and introduce weight noise ensembling to all models, improving inductive generalization and ensuring prediction stability and reliability.

We demonstrate that our approach can be applied to both graph edges and subgraph detection, enabling it to capture both local and global patterns in the graph. We evaluate \name on two large-scale, real-world graph datasets: an eBay dataset containing over 40 million transactions from 40 days for collusion detection and the Los Alamos National Laboratory (LANL) network event dataset \cite{Kent2015AuthenticationGA} consisting of 1.6 billion authentication events over 58 days for intrusion detection. These datasets present significant challenges due to their size, heterogeneity, label imbalance, and low frequency of malicious to benign events. Our experiments demonstrate that \name~ consistently outperforms state-of-the-art GNN models and baselines in both transductive and inductive settings. 
% \method~ outperforms the best-performing baselines on the eBay dataset by 0.086 in AUC and 0.151 in average precision for the transductive and inductive settings, respectively, and on the LANL dataset by 0.0053 and 0.0139 in AUC for the transductive and inductive settings, respectively. 
We summarize our contributions below

\begin{itemize}
\item We propose the first framework that integrates purely data-driven models with knowledge-enabled reasoning for enhanced anomaly detection on graph data. Our method, \namenospace, combines multiple GNN models operating on different graph structures with a probabilistic logical reasoning component, which encodes human knowledge.
\item We employ the Predictability, Computability, Stability (PCS) framework for veridical data science and introduce a weight noise ensembling approach to mitigate uncertainty and improve the inductive generalization ability of the framework.
\item We evaluate \name on two large real-world datasets: an eBay dataset for collusion detection and the Los Alamos National Laboratory (LANL) network event dataset for intrusion detection. Our results demonstrate that KnowGraph consistently outperforms state-of-the-art baselines in both transductive and inductive settings with as few as three rules.
\end{itemize}

% \bo{add a contribution list}

\section{Related Work}

\paragraph{Graph Neural Networks}

Graph neural networks (GNNs) have become a cornerstone in machine learning for graph data, effectively encapsulating local graph structures and feature information by transforming and aggregating representations from neighbors. Common architectures include graph convolutional networks (GCN) \cite{kipf2017semisupervised}, which uses convolutions directly on graph structures to capture dependencies, GraphSAGE \cite{hamilton2018inductive}, which learns inductive representations through sampling and aggregating neighborhood information, and GAT \cite{veličković2018deep}, which introduces an attention mechanism to aggregate neighborhood information. To represent long-range dependencies in disassortative graphs, Geom-GCN \cite{pei2020geomgcn} introduces a geometric aggregation scheme that enhances the convolution operation by leveraging the continuous space underlying the graph. GraphSAINT \cite{zeng2020graphsaint} proposes a graph sampling-based training method that allows for efficient and scalable learning on large graphs by iteratively sampling small subgraphs and performing GNN computations on them. DR-GCN \cite{Shi2020MultiClassIG} employs a class-conditioned adversarial network to mitigate bias in node classification tasks with imbalanced class distributions. More recently, Graphormer \cite{ying2021transformers} proposes a graph transformer model employing attention mechanisms to leverage structural information well. 

\paragraph{Generalization of GNNs}

While some standard GNN architectures \cite{hamilton2018inductive, zeng2020graphsaint} have been proposed for inductive settings, generalization remains challenging for graphs with class imbalance or diverse structures. Standard techniques such as resampling \cite{Chawla2002SMOTESM, He2009LearningFI} or reweighting \cite{Lin2017FocalLF, Huang2016LearningDR} are often sensitive to overfitting on minority classes and less effective on shifts to new graphs. Instead, many techniques aim to learn more robust and generalizable graph representations through data augmentation \cite{Zhao2020DataAF,Han2022GMixupGD} or large-scale pretraining \cite{You2020GraphCL,You2021GraphCL,Hou2022GraphMAESM}. In addition, GNN capacity can be enhanced by model combination techniques such as model ensemble \cite{sun2021adagcn,wei2023gnnensemble}, where the predictions of multiple models are combined to improve performance. Mixture-of-experts (MoE) \cite{Jacobs1991AdaptiveMO,Eigen2013LearningFR, Shazeer2017OutrageouslyLN} is a similar technique where the problem space is divided by routing inputs to specialized experts. Recent work adopts MoE for GNNs to address the class imbalance issue \cite{hu2021graph,wang2023graph,zeng2023mixture}. We also aim to address class imbalance and inductive generalization but leverage many task-specific GNNs organized through more deliberate domain knowledge.

To evaluate the generalization and uncertainty of GNNs in a principled manner, we draw upon the Predictability, Computability, and Stability (PCS) framework \cite{Yu2020VeridicalDS,Yu2024Book}. PCS emphasizes three key principles: predictability, which serves as a reality check for models; computability, which considers the feasibility and scalability of methods; and stability, which assesses the consistency of results under perturbations to data and models. In this work, we focus on the stability aspect and propose to incorporate weight noise ensembling \cite{Blundell2015WeightUI} into \name to mitigate uncertainty and improve generalization to out-of-distribution graphs.

% \begin{figure}[t]
%     \centering
%     \includegraphics[width=0.5\textwidth]{figures/lateral_movement.png}
%     \caption{A simplified version of an Advanced Persistent Threat (APT) campaign displaying the lateral
% movement cycle after initial compromise and before full domain ownership. Figure taken from \cite{Bowman2020DetectingLM}. {\color{red}make our version, more generic on anomaly detection on graph data.}}
%     \label{fig:enter-label}
% \end{figure}

\paragraph{Graph anomaly detection}

Graph anomaly detection has become an increasingly important research area, with applications spanning various domains such as network security, financial fraud detection, and social network analysis. The goal of graph anomaly detection is to identify abnormal nodes, edges, or subgraphs within a graph-structured dataset that deviates from the expected patterns or behaviors. In this paper, we focus on two representative settings: lateral movement detection and collusion detection.

In network security, graph anomaly detection is crucial in detecting and mitigating lateral movement, a key stage in the MITRE ATT\&CK framework \cite{MITRE2020}. Lateral movement refers to the propagation of malware through a network to compromise new systems in search of a target, often involving pivoting through multiple systems and accounts using legitimate credentials or malware. Research on mitigating lateral movement in computer networks generally follows three main approaches: enhancing security policies, detecting malicious lateral movement, and developing forensic methods for post-attack analysis and remediation \cite{freitas2020d2m, Hagberg2014ConnectedCA, Dunagan2009HeatrayCI, Liu2019Log2vecAH, hamilton2018inductive, Purvine2016AGI, Siadati2017DetectingSA, Ho2021HopperMA, king2023euler}. While proactive security measures can help reduce the attack surface, they cannot eliminate all potential paths an attacker might exploit. Many lateral movement detection techniques represent internal network logins as a machine-to-machine graph and employ rule-based or machine learning algorithms to identify suspicious patterns \cite{Bohara2017AnUM, Ho2021HopperMA, Kent2015AuthenticationGA}. However, these methods often struggle with scalability, generate excessive false alarms, or fail to detect attacks that do not match predefined signatures. Recent approaches improve detection by leveraging the structural nature of network data \cite{Bowman2020DetectingLM, king2023euler}, formulating the problem as a temporal graph link prediction task. These methods aim to identify edges with low likelihood scores correlated with anomalous connections indicative of lateral movement. However, challenges remain in generalizing these models to new networks in inductive settings.

In the domain of financial fraud detection, graph anomaly detection has also gained significant attention due to the economic cost and prevalence of fraud in various settings, such as social networks \cite{Breuer2020FriendOF}, online payment systems \cite{Zhong2020FinancialDD}, and online marketplace platforms \cite{Cao2017HitFraudAB, Liu2018HeterogeneousGN, Rao2020xFraudEF}. Early explorations focused on rule-based methods \cite{Cohen1995FastER} and association rules \cite{Brause1999NeuralDM, Rosset1999DiscoveryOF}, but these approaches often fail to adapt to evolving fraudulent behaviors over time. With the availability of large-scale transaction data, data-driven and learning-based methods have gained popularity, including SVM-based ensemble strategies \cite{Wang2012AHE}, graph-mining-based approaches \cite{Tian2015CrowdFD, Tseng2015FrauDetectorAG}, convolutional neural networks (CNNs) \cite{Fu2016CreditCF}, and recurrent neural networks (RNNs) for sequence-based fraud detection \cite{Jurgovsky2018SequenceCF, Wang2017SessionBasedFD, Zhang2018SequentialBD}.

More recently, GNNs have been applied to various fraud detection problems, leveraging the structural nature of transaction data \cite{Li2019SpamRD, Wang2019FdGarsFD, Weber2019AntiMoneyLI, Liu2018HeterogeneousGN, Ma2018GraphRADA}. Some works have also combined homogeneous and heterogeneous graphs to help information propagation through various types of nodes and edges \cite{Li2019SpamRD, Rao2020xFraudEF}. Despite these advancements, our proposed setting of detecting collusive fraud, where multiple adversaries conspire for illegal financial gain, remains particularly challenging due to the limited normal transaction history of colluders, severe label imbalance, and the diverse range of collusion mechanisms employed by fraudsters.

\paragraph{Knowledge-based logical reasoning}

Machine learning models are often purely data-driven and cannot directly use human knowledge to improve performance. Integrating human knowledge, such as relationships between classes, has been shown to improve ImageNet \cite{5206848} classification accuracy \cite{Deng2014LargeScaleOC} or adversarial robustness \cite{gürel2022knowledge} for deep neural networks. In addition, Bayesian logic programs \cite{Kersting2001TowardsCI}, relational Markov networks \cite{Getoor2007RelationalMN}, Bayesian networks \cite{Sebastiani1998BayesianN}, and Markov logic networks \cite{Richardson2006MarkovLN} have also been adopted for logical reasoning. Recent work has combined these methods into learning-reasoning frameworks \cite{gürel2022knowledge,yang2022improving,zhang2022care} that organize the predictions of various classifiers with knowledge rules whose weights are learned with a logical inference model. However, these works only consider simple models and tasks, such as small classifiers and datasets. The scalability of learning-reasoning frameworks, especially on large-scale graph data, has not been explored.

\section{Preliminaries}

\paragraph{Graph Neural Networks}

Given a graph $G = (V, E, Y)$, where $V$ is the set of nodes, $E$ is the set of edges, and $Y$ is the set of labels associated with the graph elements (nodes, edges, or the entire graph), the overall task of graph neural networks (GNNs) is to learn a mapping $f: G \rightarrow Y$ that predicts the labels of the graph elements based on their local neighborhood structure. To achieve this, GNNs learn representations of the graph elements by repeatedly performing neighborhood aggregation or message passing across multiple layers. The learned representations can then be used for various graph-related tasks, such as node classification ($f: V \rightarrow Y$), edge prediction ($f: E \rightarrow Y$), or graph classification ($f: G \rightarrow Y$).

Let $\mathbf{x}_j \in \mathbb{R}^{F_0}$ be the input feature vector of node $j \in V$, $h_j^{(\ell)} \in \mathbb{R}^{H}$ be the representations or embeddings of node $j$ learned by layer $\ell$ ($\ell \geq 1$), and $e_{(k,j)} \in \mathbb{R}^{F_2}$ be optional features of the edge $(k, j)$ from nodes $k$ to $j$. The message passing procedure that produces the embeddings of node $j$ via the $\ell$-th GNN layer can be described as follows:

\begin{equation}
h_j^{(\ell)} = \gamma^{(\ell)} \left( \bigoplus_{k \in \mathcal{N}(j)} \psi^{(\ell)} \left( h_j^{(\ell-1)}, h_k^{(\ell-1)}, e_{(k,j)} \right) \right),
\end{equation}

Here, $t_i^{(0)}$ is set to $x_j$, $\mathcal{N}(j)$ is the neighborhood of node $j$, and $\psi(\cdot)$ is a function that extracts a message for neighborhood aggregation, which summarizes the information of the nodes $j$ and $k$, as well as the optional edge features $e_{(k,j)}$ if available. $\bigoplus(\cdot)$ denotes a permutation-invariant function (e.g., mean or max) to aggregate incoming messages, and $\gamma(\cdot)$ is a function that produces updated embeddings of node $j$ by combining node $j$'s embeddings with aggregated messages. With multi-layer GNNs, the embeddings of a node learned via local message passing capture information from its $k$-hop neighborhood.

The final output of a GNN model, denoted as $t_i(\cdot)$, is obtained from the last layer's embeddings $t_i^{(L)}$, where $L$ is the total number of layers in the GNN. The model's prediction confidence for node $j$ is represented by $z_j$, which is derived from $t_i$ using a separate output function (e.g., a softmax layer). 

% \begin{figure*}[t]
%     \centering
%     \includegraphics[width=\textwidth]{figures/gnn-sensing-reasoning_fig1.pdf}
%     \vspace{-1.7in}
%     \label{fig:enter-label}
% \end{figure*}

\begin{figure*}[t]
    \centering
\includegraphics[width=\textwidth]{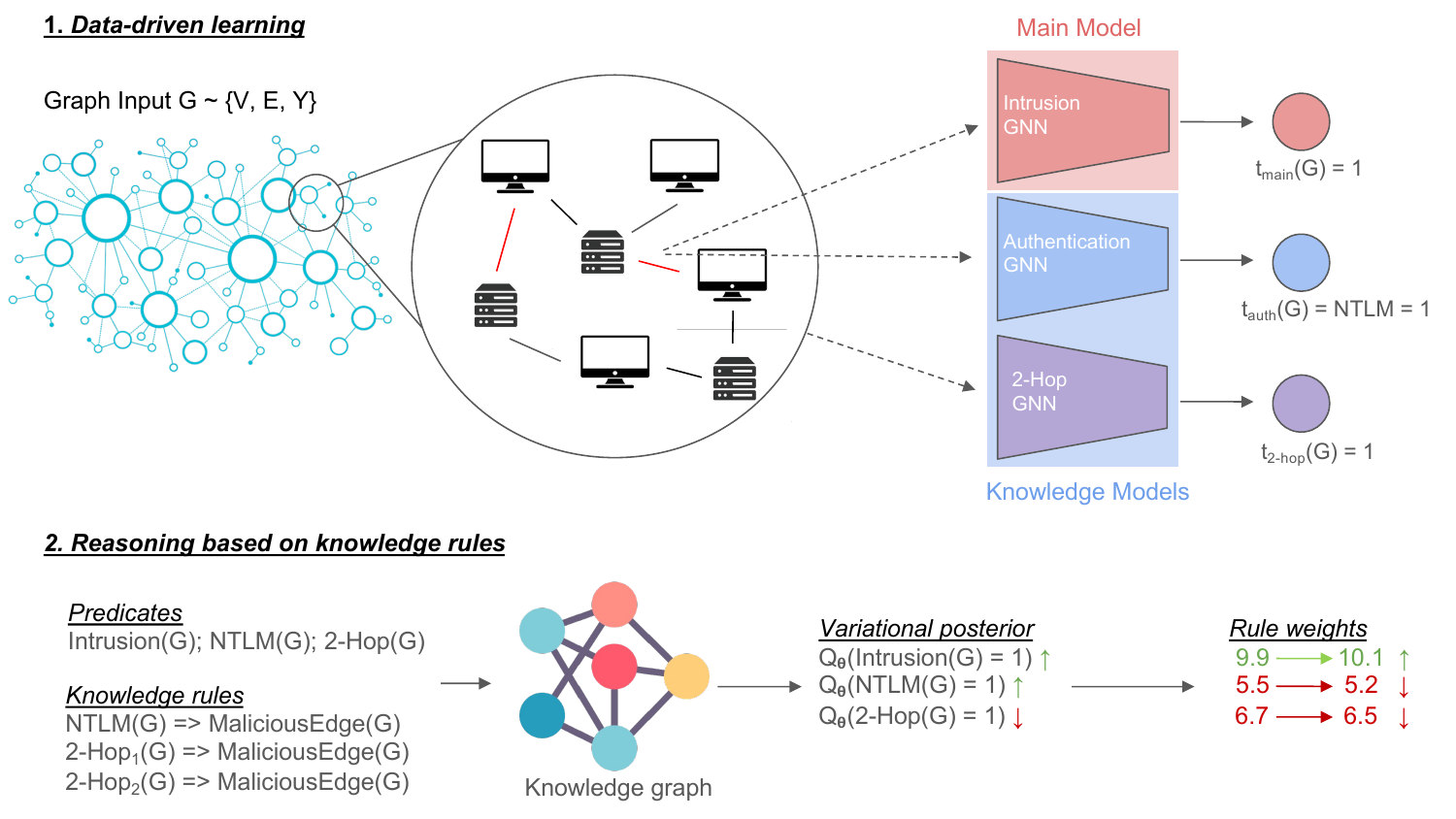}
    % \vspace{-2.7in}
    \caption{An overview of the learning and reasoning components of \method. \method~consists of a learning component composed of a main GNN model trained on the overall task and multiple knowledge GNN models trained on separate objectives, such as predicting relevant sub-attributes. The reasoning component performs logical reasoning based on the outputs of each model, which is organized based on domain knowledge rules. These rules are assigned weights, modeled by a learned scalable reasoning model parameterized by $\theta$, which explicitly ensures that the final predictions comply with the domain knowledge rules, improving reliability.}
    \label{fig:pipeline}
\end{figure*}

\paragraph{Markov Logic Networks} MLNs combine first-order logic and probabilistic graphical models to enable probabilistic reasoning over knowledge bases. An MLN is essentially a set of weighted first-order logic formulas, where each formula is associated with a weight that reflects its importance or confidence.

In the context of MLNs, the relationships between entities are represented using predicates. A predicate $p(\cdot)$ is a logical function that maps tuples of variables to binary truth values. Given a set of variables $\mathcal{V} = \{v_1,\dots,v_N\}$, where each $v_i$ represents a logical constant (e.g., an entity or an attribute), a predicate is defined as:

\begin{equation}
p(\cdot): \mathcal{V} \times \ldots \times \mathcal{V} \to \{0,1\}.
\end{equation}

First-order logic formulas in MLNs combine predicates using logical connectives (e.g., $\land$, $\lor$, $\lnot$). A formula $f(\cdot)$ is defined over a set of predicates and maps tuples of variables to binary truth values:

\begin{equation}
f(\cdot): \mathcal{V} \times \ldots \times \mathcal{V} \to \{0, 1\}.
\end{equation}

Each formula $f_i$ in an MLN is associated with a weight $w_i \in \mathbb{R}$, which represents the confidence or importance of that formula. 
% The higher the weight, the more likely the formula is to hold true.

Given a set of observed predicates (evidence) $\mathcal{E}$, an MLN defines a joint probability distribution over all possible worlds (i.e., possible assignments to all predicates):

\begin{equation}
P_w(\mathcal{X} = x | \mathcal{E}) = \frac{1}{Z(w)} \exp \left(\sum_{i=1}^{M} w_i n_i(x)\right),
\end{equation}

where $\mathcal{X}$ is the set of all predicates, $x$ is a possible world (one possible assignment for the predicates), $M$ is the number of formulas in the MLN, $n_i(x)$ represents true/false value of formula $f_i$ in $x$, and $Z(w)$ is the partition function.

Inference in MLNs involves computing the probabilities of specific predicates or formulas given the observed evidence. Various techniques, such as Markov chain Monte Carlo (MCMC) methods or lifted inference algorithms, can be used.

In the context of integrating knowledge into graph neural networks, MLNs can represent domain knowledge and perform probabilistic reasoning over the predictions of multiple GNN models. The models' outputs can be treated as observed predicates, and the MLN can be used to infer the most likely overall predictions based on the defined logical formulas and their associated weights.

\section{\name}
% In this section, we first describe an overview of our proposed sensing-reasoning framework \method. Then we describe the sensing and reasoning components of \method.

\subsection{Overview}

To effectively integrate domain knowledge into data-driven graph neural networks (GNNs), we propose \method. In particular, \method~consists of two components: a \textit{learning} component and a \textit{reasoning} component. Specifically, the \textit{learning} component comprises a main model and several knowledge models. The main model focuses on the main function (e.g., malicious edge detection in intrusion detection), delivering multi-class predictions based on graph inputs. The knowledge models take the same graph as input to predict specific semantic entities (e.g., authentication type of an edge). The internal relationships among these models are represented by expert-designed knowledge rules (e.g., a malicious edge indicates a specific authentication type), which are embedded in a reasoning component for logical inference.
 % These entities can be \textit{directly incorporated into the knowledge rules} when available or labeled. 
For example, in an edge-level intrusion detection task, the main model is trained to predict whether an edge is malicious, and a knowledge model is trained to predict the authentication type of the edge. The outputs of these models are then sent to a reasoning component to check whether they satisfy the domain knowledge rules such as ``malicious edges indicate an NTLM \cite{Microsoft2010NTLM} authentication type''. During inference, if the knowledge is violated, the prediction of the main model will be automatically corrected, leading to a more resilient final prediction. 
% aided by the inclusion of expert domain knowledge such as ``malicious edges indicate a NTLM \cite{Microsoft2010NTLM} authentication type'', and a knowledge model is trained to predict the authentication type of an edge. 
In \namenospace, every model in the learning component is a trained GNN model, distinguishing \name from some previous frameworks \cite{gürel2022knowledge,yang2022improving,zhang2022care} that rely on simpler deep neural network (DNN) classifiers.

Concretely, the knowledge rules are formalized as first-order logic rules, such as ``\texttt{IsNTLM(x)} $\implies$ \texttt{IsMalicious(x)}'' within the \textit{reasoning} component, which learns a weight for each rule from its direct usefulness. The \textit{reasoning} component itself is implemented using probabilistic graphical models, such as a Markov Logic Network (MLN) \cite{Richardson2006MarkovLN}. Due to the computationally intensive nature of MLN inference from the exponential complexity of constructing the ground Markov network, we use a scalable variational inference method \cite{zhang2022care} to learn a GNN to embed the knowledge rules. We utilize the Expectation-Maximization (EM) algorithm that iteratively enhances the GCN's accuracy in learning the GCN model weights (E step) and refines the weighting of knowledge rules (M step) within the latent MLN. During inference, the learned reasoning GCN component ensures that the prediction outputs of the main and knowledge models satisfy the defined domain knowledge rules and achieve resilient final predictions.

\begin{figure*}[t]
    \centering
    \includegraphics[width=1.0\textwidth]{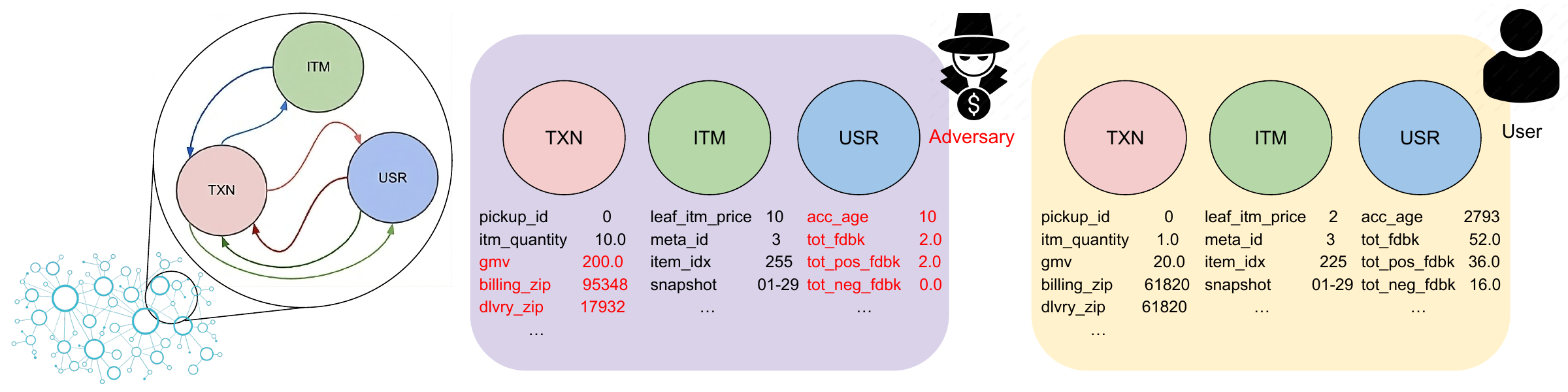}
    \caption{An illustration of the graph from real eBay marketplace data containing the core entities of a transaction, which are represented as nodes in a knowledge graph consisting of nodes of transactions (TXN), users (USR), and items (ITM). Examples of collusive and benign transactions are shown. Example node attributes are listed below the entity, such as gross merchandise value (gmv), account age (acc$\_$age), and total feedback (tot$\_$fdbk). Domain knowledge suggests that collusive transactions typically involve discrepancies between billing and delivery zip codes and are associated with users who have minimal feedback and newer accounts. This understanding has led to the formulation of rules such as ``$[\texttt{feedback amt} < a] \land [(\texttt{seller\_age} < b \lor \texttt{buyer\_age} < c)] \Rightarrow \texttt{collusion}$'', which help refine the main model's predictions based on these indicators.  %\bo{write full name for TXN, ITM. add name for collusion and benign. left fig could be larger containing multiple such subgraphs?}
    }
    \label{fig:ebay-figure}
\end{figure*}

\subsection{Learning with knowledge-enabled reasoning on graph data}

In the \textit{learning} component of \method, we train a set of GNN models to predict the main task and knowledge models for assistant tasks. The knowledge models are conceptualized as predicates within this framework, outputting binary predictions for each knowledge model.

Formally, we denote the output of the i-\textit{th} model by $t_i(\cdot)$ as $t_i$, with $z_i$ representing the confidence level of its prediction. For a given input graph $G = (V, E, Y)$ with nodes $V$, edges $E$, and labels $Y$, the GNN model's prediction is denoted as $t_i(G)$. Upon receiving input $G$ and the assorted model predictions $t_i(G)$, these predictions are interlinked through their logical interrelations and a Markov Logic Network (MLN) \cite{Richardson2006MarkovLN}, enabling the \textit{reasoning} capability in \method.

Specifically, \method~involves a primary model alongside multiple knowledge models $t_i(G)$, serving as predicates within the MLN framework. Logical connections between these predicates are established to formulate different logical expressions. Assuming $L$ models in total, an MLN defines a joint probability distribution over the pre-defined logical expressions (i.e., knowledge rules), which can be expressed as follows:

\begin{equation}
\label{eq:joint_p}
P_w(t_1,...,t_L) = \frac{1}{Z(w)} \exp \left( \sum_{f \in \mathcal{F}} w_f f(t_1,...,t_L)\right),
\end{equation}
with $Z(w)$ symbolizing the partition function, summing across all predicate assignments.

\method's reasoning component manages logic formulas articulated as first-order logic rules. Following \cite{zhang2022care}, we consider three types of logic rules:

\begin{itemize}
    \item Attribute rule ($t_i \implies t_j \lor t_j \lor... $): This rule leverages specific attributes associated with prediction classes to formulate knowledge-based rules.
    \item Hierarchy rule ($t_i \implies t_j$): Reflecting the hierarchical nature among classes, this rule aids in constructing logical expressions like $f(t_i, t_j) = \lnot t_i \lor t_j$.
    \item Exclusion rule ($t_i \oplus t_j$): This rule addresses the inherent exclusivity among some class predictions, ensuring that an entity cannot simultaneously belong to mutually exclusive classes.
\end{itemize}
After designing the models and rules, the final step of \method~ is to learn and assign a weight for each rule to reflect the impact of their prediction confidence $z_i$ for each model $t_i(\cdot)$. To achieve this, we utilize the logarithm of the odds ratio, $\log [z_i/(1-z_i)]$, as the weight for model $t_i$.

\subsection{Scalable reasoning with a GCN}
% \xiaojun{Suggest trimming this section - it's not our contribution?}
% \bo{we can illustrate the challenges on graph data, and then explain the method in detail}
% In this section, we detail the approach for approximating the inference process in our Markov Logic Network (MLN) through variational inference, facilitated by Graph Convolutional Networks (GCN). Our objective centers on optimizing the knowledge rule weights within the MLN to maximize the log-likelihood of observed predicates. 

To reduce the computational complexity of training the MLN, we employ variational inference \cite{zhang2022care} to optimize the variational evidence lower bound (ELBO) of the data log-likelihood. This approach is motivated by the intractability of directly optimizing the joint distribution $P_{w}(\mathcal{O}, \mathcal{U})$, which requires computing the partition function $Z(w)$ and integrating over all observed predicates $\mathcal{O}$ and unobserved predicates $\mathcal{U}$. The ELBO is formulated as follows:
\begin{equation}
    \begin{split}
    \log P_{w}(\mathcal{O}) \ge \mathcal{L}_{\mathrm{ELBO}}(Q_{\theta}, P_{w}) &=  \mathbb{E}_{Q_{\theta}(\mathcal{U} | \mathcal{O})}\left[\log P_{w}(\mathcal{O}, \mathcal{U})\right] \\
    & - \mathbb{E}_{Q_{\theta}(\mathcal{U} | \mathcal{O})}\left[\log Q_{\theta}(\mathcal{U} | \mathcal{O})\right],
    \end{split}
\end{equation}
where $Q_{\theta}(\mathcal{U}|\mathcal{O})$ is the variational posterior distribution. The representation of model outputs and knowledge rules as a graph motivates the use of Graph Convolutional Networks (GCNs) for encoding $Q_{\theta}(\cdot)$.

We adopt a variational EM algorithm to refine the ELBO and learn the MLN weights $w$. In the E-step, the GCN parameters $Q_{\theta}$ are updated to minimize the KL divergence between $Q_{\theta}(\pT)$ and $P_w(\pT)$, where $\pT={t_1, t_2, ...,t_L}$ are the model outputs. The optimization objective is enhanced with a supervised negative log-likelihood term $\mathcal{L}_{\text{supervised}}$ to leverage available label information during training:
\begin{equation}
    \label{eq:new-elbo}
    \begin{split}
    \mathcal{L}_{\mathrm{ELBO}}(Q_{\theta}, P_{w}) := \mathbb{E}_{Q_{\theta}(\pT)}[\log P_{w}(\pT)] - &  \mathbb{E}_{Q_{\theta}(\pT)}[\log Q_{\theta}(\pT)].
    \end{split}
\end{equation}
where $\eta$ is a hyperparameter balancing the importance of the ELBO and the supervised term. This approach ensures that the class embedding vectors $\vec{\mu}$ are refined during the optimization of the GCN, enhancing the model's expressiveness.

Conversely, the M-step fixes $Q_{\theta}$ while updating the weights $w$. Since directly integrating over all variables is intractable, we follow \cite{zhang2022care} and optimize the pseudo-likelihood~\cite{Besag1977EfficiencyOP} instead, which is formulated as:
\begin{equation}
\begin{aligned}
P_{w}^*(t_1,..., t_L) := \prod_{i=1}^{L} P_{w}\left(t_{i} | MB\left(t_{i}\right)\right), 
\end{aligned}
\end{equation}
where $f\left(t_{i}=0\right)$ and $f\left(t_{i}=1\right)$ reflect the formula $f$'s ground truth values under the hypothetical scenarios of $t_i$ being 0 or 1, respectively, with other variables held constant.

This strategy addresses the intractability of directly optimizing the complex log-likelihood by maximizing the expectation of the pseudo-log-likelihood, which is estimated through multiple samplings from the variational distribution $Q_{\theta}$.

For a more detailed description of the variational inference procedure, including the derivation of the equations and the specific formulations of the ELBO terms, please refer to \cite{zhang2022care,ng2012algorithm}.

\input{tables/lanl_descriptions}

\subsection{Uncertainty mitigation with PCS}

To further improve inductive generalization, we draw upon the Predictability, Computability, and Stability (PCS) framework \cite{Yu2020VeridicalDS,Yu2024Book}, which is a comprehensive approach to ensure the reliability, reproducibility, and transparency of data-driven results when employing complex modeling techniques. The PCS framework emphasizes three key principles: predictability, which serves as a reality check for models; computability, which considers the feasibility and scalability of methods; and stability, which assesses the consistency of results under perturbations to data and models. When applying \name to critical real-world tasks, uncertainty quantification helps to assess and mitigate the uncertainty associated with our results. Adding uncertainty information to \name can be valuable for decision-making, as it allows us to identify cases where the model is less confident and may require additional investigation or human intervention.

In particular, the stability principle in PCS emphasizes the importance of assessing the stability of data results with respect to data and model perturbations. We propose incorporating a weight noise ensembling technique
% \cite{Blundell2015WeightUI} 
into the stability principle of the PCS framework. Weight noise ensembling introduces random perturbations to the weights of the neural network during training and inference, allowing us to capture the uncertainty in the model's parameters by considering a distribution over the weights rather than a single point estimate. This approach acts as a regularization technique, reducing overfitting in our setting with extreme label imbalance and promoting the learning of robust features that can generalize to new graphs.

To integrate weight noise ensembling into our pipeline within the PCS framework, we modify the statistical learning and reasoning components of \namenospace. Within the learning component, we train multiple instances of the GNN model, each with different random perturbations added to the weights. These perturbations are sampled from a predetermined probability distribution (e.g., Gaussian distribution with mean 0 and a specified variance). In the reasoning component, we perform inference using each perturbed model and collect the predictions from each instance. We then compute summary statistics (e.g., mean and variance) of the ensemble predictions to mitigate the uncertainty associated with each collusion detection result, allowing us to make more robust updates to the learned rule weights and GCN posterior.

\textit{Discussion.} Conceptually, KnowGraph has several intuitive advantages as a framework for graph-based anomaly detection: (1) \textit{Flexibility}: KnowGraph improves performance with even a small number of well-designed rules. The usefulness of each rule is encoded in its learned weight, allowing the framework to default to the original base model prediction if the rule is not helpful. (2) \textit{Scalability}: The learning component trains knowledge models in parallel, scaling linearly with the number of models. The reasoning component's complexity depends on the number of rules, where each rule adds only a few nodes and edges to the reasoning component, allowing for easy expansion of knowledge models and rules without significant computational overhead. (3) \textit{Adaptability}: By combining data-driven learning with knowledge-based reasoning, KnowGraph demonstrates strong generalization capabilities. This hybrid approach allows the framework to adapt to new patterns and evolving scenarios, making it particularly suited for dynamic environments where attack strategies may change over time. Both new knowledge models and rules can also be added, only requiring the reasoning component to be retrained. 

% During evaluation, we quantify the uncertainty in the model's predictions using two metrics: the Brier score and the Expected Calibration Error (ECE). The Brier score measures the accuracy of probabilistic predictions, taking into account both the predicted probability and the actual outcome. It is defined as:

% \begin{equation}
% BS = \frac{1}{N} \sum_{i=1}^{N} (p_i - y_i)^2
% \end{equation}

% where $N$ is the number of instances, $p_i$ is the predicted probability of the positive class for instance $i$, and $y_i$ is the true label (0 or 1) for instance $i$. A lower Brier score indicates better calibrated and more accurate predictions.

% The Expected Calibration Error (ECE) measures the difference between the predicted probabilities and the actual empirical frequencies of the outcomes, averaged over bins of predicted probabilities. It is defined as:

% \begin{equation}
% ECE = \sum_{m=1}^{M} \frac{|B_m|}{N} |\text{acc}(B_m) - \text{conf}(B_m)|
% \end{equation}

% where $M$ is the number of bins, $B_m$ is the set of instances whose predicted probabilities fall into the $m$-th bin, $\text{acc}(B_m)$ is the accuracy of the predictions in bin $B_m$, and $\text{conf}(B_m)$ is the average predicted probability in bin $B_m$. A lower ECE indicates better calibrated predictions that align with the actual empirical frequencies. We observe adding weight ensembling improves calibration for both ECE and Brier score.

\section{Experiments}

\subsection{Intrusion detection on LANL}

In this section, we describe our results on intrusion detection on the public Los Alamos National Labs (LANL)~\cite{kent2016cyber} dataset.

\paragraph{Dataset} We conduct experiments on a open-sourced intrusion detection dataset\footnote{Detailed dataset information can be found at \url{https://csr.lanl.gov/data/cyber1/}.} from Los Alamos National Labs (LANL)~\cite{kent2016cyber}. The LANL dataset contains a 58-day log within their internal computer network, among which the malicious authentication events are identified. The dataset comprises 12,425 users, 17,684 computers, and 1.6B authentication events. We model the dataset as a graph so that a node represents a server in the network and an edge represents an authentication event (e.g. log-in / log-out) from one server to another. We will remove the repeated edges in the graph so that the overall number of edges is 45M. Our goal is to detect whether an edge is malicious or not.

\paragraph{Labels} Each edge in the graph has a ground-truth label by the dataset, indicating whether the authentication event is malicious or not. Only 518 malicious edges are among the 45M edges, making the dataset highly unbalanced.

\input{tables/lanl-overall}

\paragraph{Settings and baselines}
We mainly follow the data pre-processing procedure as in the Euler~\cite{king2023euler} work. We use the training and validation set that consists of all the information before the first anomalous edge appears, with 5\% of them being the validation set and the rest being the training set. We split the dataset into multiple graphs so that each graph consists of the edges within a time window of 1,800 seconds. After training the model, we will evaluate it on the test set, which consists of the edges after the first anomalous edge appears. To compare the detection performance, we will evaluate the AUC, Average Precision (AP), and the True Positive rate at a certain False Positive rate (e.g., TP@0.5FP denotes the true positive rate when the false positive rate is 0.5\%).

We adopt two baselines. The first baseline is the Euler~\cite{king2023euler} work, where the authors use graph neural networks (with or without a recurrent neural network header) to detect the malicious edges. The inputs to the model include the node and edges within a time window. The GNN processes the information and returns a value for each edge, indicating the probability that it is malicious. The model will be trained with negative sampling. In each training step, the model will be trained to give a low malicious probability for the existing edges and a high malicious probability for a randomly sampled set of non-connected node pairs.
% \bo{control FP 0.5\%, 1\%，2\% and report TP for methods}

The second baseline uses link prediction with an Enclosing Graph (EncG)~\cite{zhang2018link}. EncG extracts the K-hop enclosing subgraph for each edge and trains a subgraph classification model to determine whether the corresponding edge is malicious. We use $K=2$ considering the tradeoff between efficiency and effectiveness.

\begin{figure}
    \centering
    \includegraphics[width=0.9\linewidth]{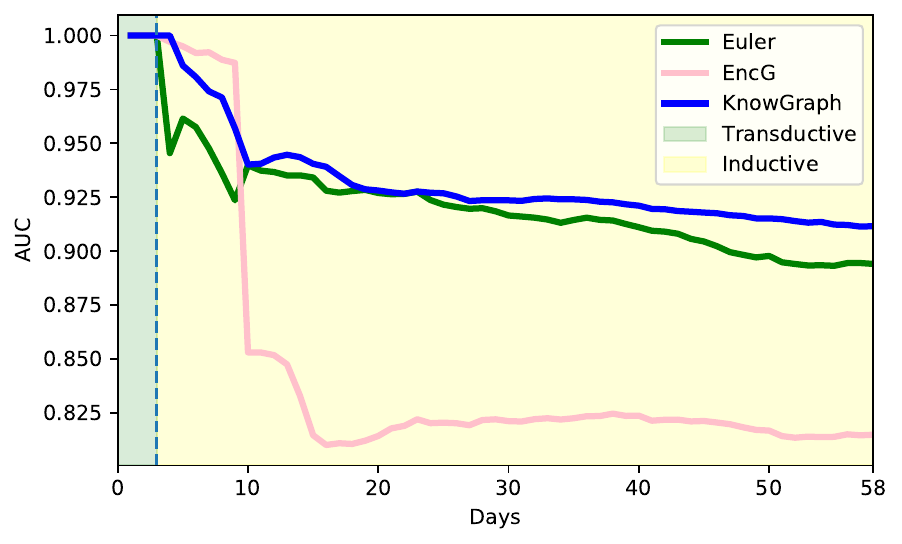}
    \caption{Detection AUC of baselines and \name given different time shifts on the LANL dataset. In the challenging inductive setting (yellow), the baseline performance drops significantly while \name still maintains high detection performance.}
    \label{fig:lanl-auc}
\end{figure}

\paragraph{Knowledge models}
We have three models for this task. First, we have the \texttt{main} model to determine whether the edge is malicious, with the same model architecture as in the Euler work. Second, we have the \texttt{auth} model, which is also a binary edge classifier to determine whether the authentication type of the model is NTLM or not. Finally, we have a \texttt{EncG} model that follows the same methodology in the EncG~\cite{zhang2018link} work to judge the edge maliciousness based on the enclosing graphs.
% also determines whether an edge is malicious or not, but only with the information of a 2-hop enclosing graph. This model follows the idea in \citep{zhang2018link} which extracts the a K-hop enclosing subgraph for each edge and performs a subgraph classification.   

\paragraph{Knowledge rules}
We designed three rules for the task, which are as follows: (1) \emph{Authentication rule}, which states that malicious authentications are likely to be in type NTLM, ``$[\texttt{main} = \texttt{Mal}] \Rightarrow [\texttt{auth} = \texttt{NTLM}]$''. This rule incorporates the human knowledge that NTLM authentication is usually a less secured authentication protocol (compared with, for example, Kerberos), and most malicious authentications are using NTLM authentication. (2) \emph{Subgraph rule 1} which states that the main model should be malicious if 2hop model predicts a malicious, ``$[\texttt{EncG} = \texttt{Mal}] \Rightarrow [\texttt{main} = \texttt{Mal}]$''. (3) \emph{Subgraph rule 2} which states that the main model should be benign if 2hop model predicts a benign, ``$[\texttt{EncG} = \texttt{Benign}] \Rightarrow [\texttt{main} = \texttt{Benign}]$''. Note that rules (2) and (3) combined indicate that the prediction outputs of the \texttt{main} model and the \texttt{2hop} model should be consistent.

% \paragraph{Main Results}
% \input{tables/lanl-overall}
% \begin{figure}
%     \centering
%     \includegraphics[width=0.9\linewidth]{figures/lanl_auc_time.pdf}
%     \caption{Detection AUC of Euler and our work on different time domain.}
%     \label{fig:lanl-auc}
% \end{figure}
% \begin{figure}
%     \centering
%     \includegraphics[width=0.9\linewidth]{figures/Euler_lanl_timedomain.pdf}
%     \includegraphics[width=0.9\linewidth]{figures/encg_lanl_timedomain.pdf}
%     \includegraphics[width=0.9\linewidth]{figures/ours_lanl_timedomain.pdf}
%     \caption{Model prediction logits on different time domain for (top) Euler, (middle) EncG, and (bottom) KnowGraph. A low prediction means that the access is considered as malicious. We can observe that all approaches have a good performance for transductive setting; for inductive setting, it is easier for \method~to separate the malicious accesses, by setting a threshold at around -1.5. By comparison, it is difficult to classify between benign and malicious accesses.}
%     \label{fig:lanl-pred}
% \end{figure}

\paragraph{Main Results}
Based on our intensive evaluations, we show the model and detection performance of the final reasoning pipeline in Table~\ref{tab:lanl-overall}.
We observe that i{n the transductive setting, all models perform well with close performance given similar training and testing data distributions. In particular, the detection performance of the baselines can achieve over 0.99 AUC in the inductive setting. By integrating the model information, we can improve the performance and achieve a close-to-perfect detection performance. However, the model performance drops to below 0.9 under the inductive learning setting with out-of-distribution data. 
% \bo{add analysis for inductive and transductive settings and emphasize that our method is much better under inductive setting and also that's our goal}
By integrating knowledge rules and reasoning, \name can better aggregate the information from the models in this OOD scenario and significantly improve detection performance. This showcases the robustness of the reasoning framework against OOD cases and matches our intuition of including human knowledge in the pipeline.

% \begin{figure}
%     \centering
%     \includegraphics[width=0.9\linewidth]{figures/lanl_auc_time.pdf}
%     \caption{Detection AUC of Euler and our work on different time domain.}
%     \label{fig:lanl-auc}
% \end{figure}
\begin{figure}
    \centering
    \includegraphics[width=0.9\linewidth]{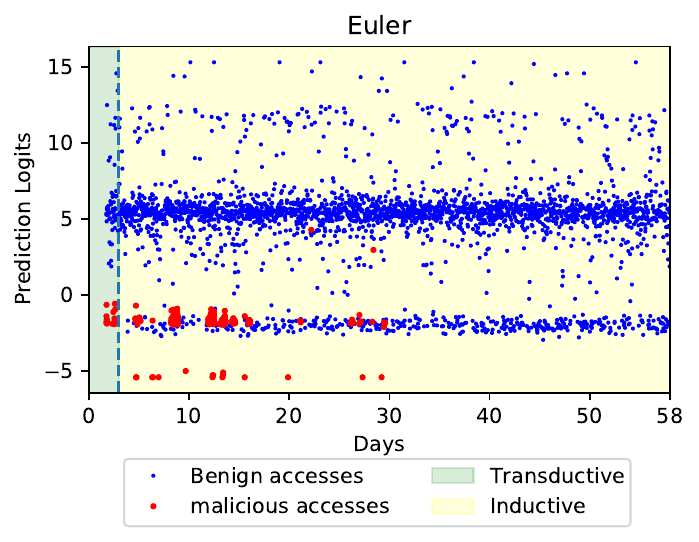}
    \includegraphics[width=0.9\linewidth]{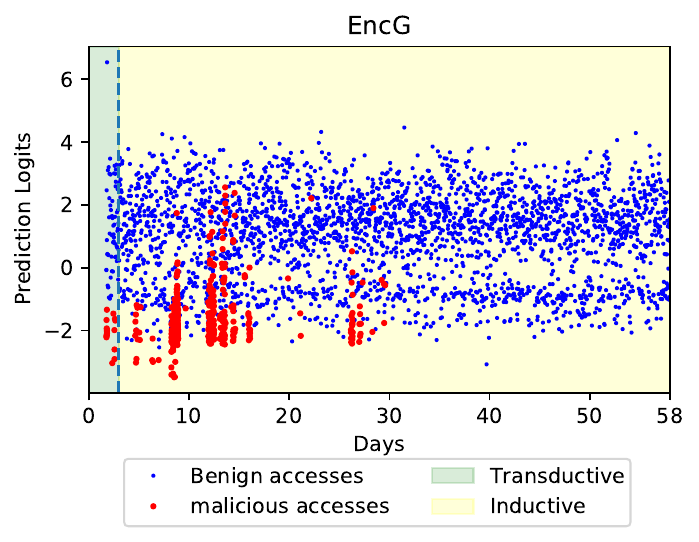}
    \includegraphics[width=0.9\linewidth]{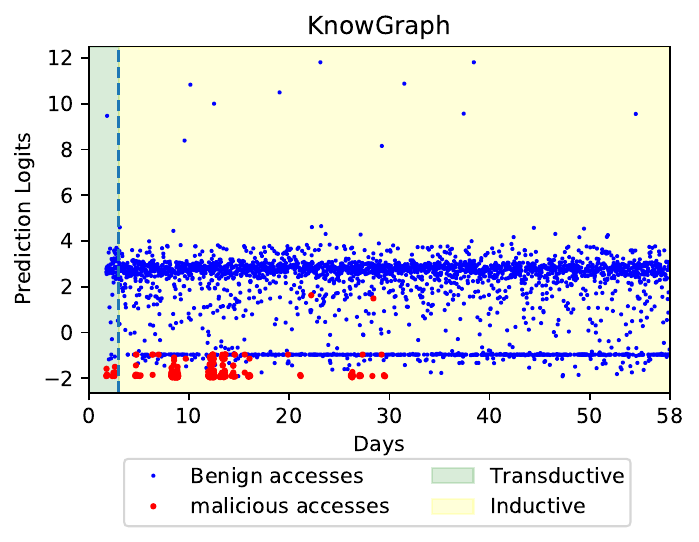}
    \caption{Model prediction logits on the different time shifts for (top) Euler, (middle) EncG, and (bottom) KnowGraph. Low prediction logits indicate that the access is considered malicious. We can observe that all approaches perform well for the transductive setting; for the inductive setting, it is easier for \method~to separate the malicious accesses by setting a threshold at around -1.5. However, it is difficult to classify benign and malicious accesses based on the prediction logits of Euler and EncG.}
    \label{fig:lanl-pred}
\end{figure}

In addition, we also show the detection AUC and prediction logits change on the time domain for both Euler and our method in Figure~\ref{fig:lanl-auc} and Figure~\ref{fig:lanl-pred}, respectively.
% Our reasoning framework shows better robustness with respect to the OOD samples.  
% \bo{provide some intuitions, such as knowledge rules are stable over time while data distributions will change and such knowledge rules are important for genrationalization and explanation}
We can observe that, as the timestamp goes to the later days, the detection AUC will gradually drop due to the OOD issues. For both Euler and EncG, many benign OOD cases are viewed as malicious by the model, which is not seen in the transductive setting. Moreover, many malicious accesses receive a low malicious score in EncG. By comparison, the distribution of benign and malicious accesses is generally similar in both transductive and inductive settings. We attribute it to the stability of the reasoning framework with human knowledge.

\input{tables/lanl-abl}

\paragraph{Ablation Studies} We show the ablation study of model performance in Table~\ref{tab:lanl-abl}. We can observe that every model has a relatively good performance (with transductive AUC larger than 0.9 and inductive AUC larger than 0.8). After combining the models with our reasoning framework, we can further improve the performance in both transductive and inductive settings.
% \bo{maybe describe the rational and analysis more clearly here}
Note that the reasoning framework is ubiquitously better than each of its components, showing its superiority over simple methods, which, for example, average the results. In addition, we observe a trend that combining better-performing models can yield better reasoning results, which shows the importance of training better models and designing the reasoning framework. 

% \paragraph{figures}
% \begin{figure}
%     \centering
%     \includegraphics[width=0.9\linewidth]{figures/lanl_auc_time.pdf}
%     \caption{LANL AUC}
%     \label{fig:lanl-auc}
% \end{figure}
% \begin{figure}
%     \centering
%     \includegraphics[width=0.9\linewidth]{figures/Euler_lanl_timedomain.pdf}
%     \includegraphics[width=0.9\linewidth]{figures/ours_lanl_timedomain.pdf}
%     \caption{Model prediction on the time domain for Euler(top) and Reasoning(bottom). Lower means more malicious.}
%     \label{fig:lanl-pred}
% \end{figure}
% Time-domain prediction: Figure~\ref{fig:lanl-auc} and Figure~\ref{fig:lanl-pred}.

% \paragraph{Ablation Study - Correlation within models}
% See Table~\ref{tab:lanl-abl-corr}
% \begin{table}
%   \centering
%   \caption{Ablation study - Correlation with \texttt{main} model vs. Performance combining with \texttt{main} model. Evaluated on the transductive setting.}
%   \label{tab:lanl-abl-corr}
%     \begin{tabular}{l|cccc}
%     \toprule
%     model & Correlation to \texttt{main} & AP & AP(+\texttt{main}) \\
%     \midrule
%     \texttt{EncG} & 0.1394 & 0.3249 & 0.5320 \\
%     \texttt{main(1-layer)} & 0.9562 & 0.0623 & 0.0521 \\
%     \texttt{main(3-layer)} & 0.7811 & 0.0332 & 0.0392 \\
%     \bottomrule
%     \end{tabular}
% \end{table}
\input{tables/ebay_descriptions}

\subsection{Collusion detection on real-world eBay marketplace dataset}

In this section, we describe the real-world eBay marketplace dataset and our collusion detection results and analysis in detail. 

\paragraph{Dataset} We use a large-scale proprietary dataset on real-world marketplace transactions from the popular online shopping website eBay, which has more than 135 million users \cite{yaguara2023ebay}. The dataset contains transactions from 40 days total, each with around 4 million transactions, collected from January to February of 2022. Each transaction consists of three entities: a seller, a buyer, and an item. These transactions and entities are organized into a knowledge graph, where each entity is a node with bidirectional edge relationships with other entities in the same transaction. The corresponding task aims to train models that predict if a transaction indicates buyer-seller collusion, a type of fraud in which buyers and sellers conspire for illegal financial gain. This fraud can involve manipulating prices or exchanging fake feedback to deceive the marketplace, leading to significant financial losses. Due to the scale of the marketplace, this has a financial cost of millions of dollars a year based on proprietary estimates, making automated detection crucial.

\vspace{-1mm}

\paragraph{Labels} Each transaction has ground-truth collusion labels from real collusion cases and additional data based on various buyer, seller, and item features. These include relevant knowledge features such as transaction zip code, item price, and shipping cost, comprising 26 features across the three entities. This is summarized in Fig.~\ref{fig:pipeline}, which contains example differences between benign and anomalous transactions. The ground-truth collusion labels include a single overall binary classification label and multiple fine-grained class or collusion subclass labels. Subclasses are based on specific instances of collusion, such as collusion from a particular group or those exhibiting specific collusive modus operandi. The number of positive labels is extremely sparse, with only around 1330 new collusion cases daily. This makes the exact ratio of positive to negative labels $1: 3269.371$, a case of extreme label imbalance.
% \bo{refer to fig3 and describe in detail?}

% \input{tables/ebay_overall}

% % \input{tables/ebay_fpr}

% \input{tables/ebay_descriptions}

\paragraph{Settings and baselines} To manage label sparsity in our data, we segment the graph into temporal subgraphs, each spanning 20 consecutive days. The initial model training is conducted on a subgraph containing the first 20 days of transactions, a portion of which is reserved as a test set for evaluation. We use additional 20-day snapshots from subsequent periods to test time-shift generalization in the inductive setting. These later snapshots vary in their degree of temporal overlap with the training snapshot; 'overlap' here refers to days that are included in both the training snapshot and testing snapshots. In the most challenging setting, the test snapshot comprises the latter 20 days and has no overlapping days with the training set. We evaluate the AUC, Average Precision (AP), and the True Positive rate at a certain False Positive rate (e.g., TP@0.5FP denotes the true positive rate when the false positive rate is 0.5\%).

The data is organized into a Heterogeneous Knowledge Graph \cite{Hogan_2021} comprising three core entities in the eBay e-commerce setting -- users, items, and transactions. Edges define transactional relationships between buyers and sellers (users) for specific items. The heterogeneous knowledge graph with GCN layers is trained on node graph features with Deep Graph Infomax (DGI) \cite{veličković2018deep} to learn label agnostic node embeddings. Inductive prediction generates heterogeneous node embeddings for the four key entities within the transactional sub-graph: transaction, buyer, seller, and item. These embeddings are combined to form a feature set for the sub-graph, which is then input into an XGBoost \cite{Chen_2016} classifier trained to perform node-level collusion classification. This configuration serves as the primary baseline and main model within \namenospace, and is also used in the GNN architecture for the knowledge models. It is also the main component in the current automated collusion detection system at eBay, and our overall goal is to improve its performance and generalization. We also compare using only domain knowledge and a simpler GCN \cite{kipf2017semisupervised} baseline trained with supervised learning, which has strong transductive performance but weaker generalization. Since we focus on the inductive setting, it is not used as a model in \namenospace.

\input{tables/ebay_overall}

% \input{tables/ebay_fpr}

\paragraph{eBay Dataset Challenges} For the transductive setting, the model has access to the entire graph during training but not the collusion labels for nodes in the test set. The model must generalize to a new snapshot with an unseen graph for the inductive setting. We control the difficulty of this setting by changing the amount of overlapping days with the training snapshot.
We observe that generalization is challenging in this context due to several reasons:
\begin{itemize}
\item The high volumes of eBay transactions (4-5 million per day) provide substantial changes to the graph structure and label distribution across graph snapshots. 
\item The noise induced by the inductive sub-graph sampling for graph inference can be significant as the entire graph is too large for tractable message passing. 
\item The extreme label imbalance coupled with collusive participants who are mostly new to the platform, often lacking prior purchasing or selling history, composes fewer anomalous links in the graph. 
\item The dataset does not model dynamic time-varying features for users and items, instead relying on static input node features for each such reference entity within a single snapshot. 
\end{itemize}

Due to these challenges, baselines cannot generalize to the inductive setting and have performance close to random guessing. However, while not as effective for the transductive setting, expert-designed knowledge rules have higher performance due to invariance to this distribution shift.

\begin{figure}[t]
    \centering
    \hspace{-4mm}\includegraphics[width=0.49\textwidth]{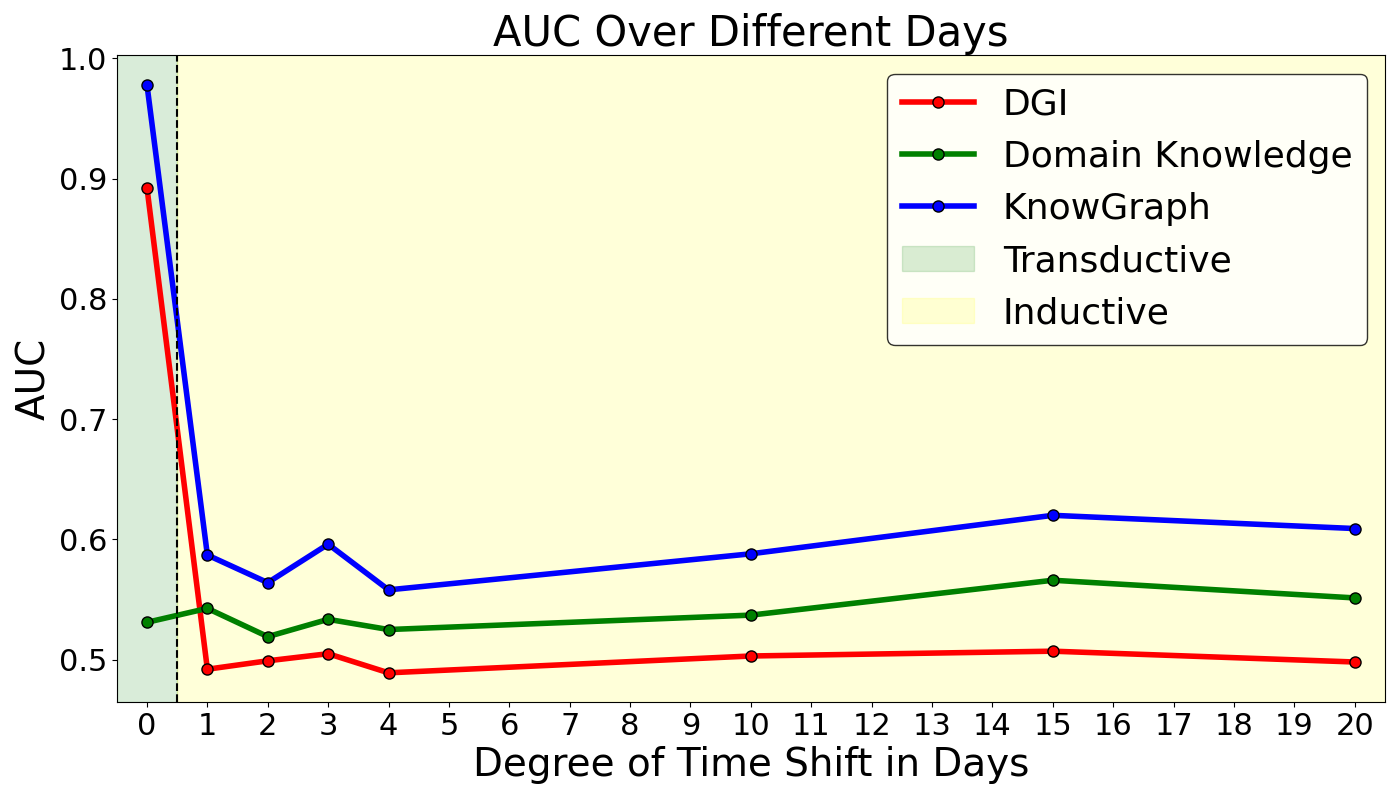}
    \caption{AUC of test graph snapshots with decreasing overlap with the training snapshot. 
    % On day 20, there is no overlap, and the setting is fully inductive. 
    \name consistently outperforms baselines.
    We observe similar performance regardless of how different the new graph is, indicating \name is robust to the magnitude of the time shifts.}
    \label{fig:auc_chart}
\end{figure}

\input{tables/ebay_rule_weights}

\input{tables/ebay_model}

% \begin{figure}[t]
%     \hspace{-4mm}\includegraphics[width=0.49\textwidth]{figures/avg_precision_chart_final.png}
%     \caption{Average Precision of test graph snapshots with decreasing overlap with the training snapshot. Similar to Fig.~\ref{fig:auc_chart} we observe the improved performance of \name over baselines consistently under different time drifts.}
%     \label{fig:avg_precision_chart}
% \end{figure}

\paragraph{Knowledge models} For the learning component of the framework, we augment the unsupervised embedding model trained on the main classification task with several knowledge models centered around more specific tasks for a total of eight models. Utilizing the fine-grained collusion labels, we train a model for each fine-grained class for four models. These labels are derived from specific instances or cases of collusion. Next, we train a second binary collusion model on upsampled data. These models are independently trained XGBoost classifiers on their respective tasks using the same unsupervised graph embeddings. Finally, we use graph features on buyer and seller account age, zip code, and price to construct two additional models. Unlike the other models, these are matrices directly derived from the graph.

\paragraph{Knowledge rules} For the reasoning component of the framework, we organize the model's predictions into logical rules. The models are predicates of the main model and are organized into the following six rules. 

\begin{itemize}
    \item \textit{Hierarchy rule} that denotes the hierarchical relation between the binary and collusion subclasses, 
    \\
    ``$[\texttt{subclass} = \texttt{collusion}]\Rightarrow [\texttt{main} = \texttt{collusion}]$.'' 
    \item \textit{Exclusion rule} that utilizes the mutually exclusive nature of fine-grained classes, formulated as ``$[\texttt{collusion}_i = \texttt{true}] \Rightarrow [(\texttt{collusion}_{0} = \texttt{false}) \land \dots \land (\texttt{collusion}_{i-1} = \texttt{false}) \land (\texttt{collusion}_{i+1} = \texttt{false}) \land \dots \land (\texttt{collusion}_n = \texttt{false})]$.'' where $i$ is the class index up to $n=4$ 
    \item \textit{Ensemble rules} that ensembles the predictions of the two models trained on the overall classification task by ensuring consistency, 
    \\
    ``$[\texttt{main} = \texttt{collusion}]  \Rightarrow [\texttt{secondary} = \texttt{collusion}]$'' and ``$[\texttt{secondary} = \texttt{collusion}] \Rightarrow [\texttt{main} = \texttt{collusion}]$.'', where secondary refers to the model trained on a larger graph with balanced sampling.
    \item \textit{Business rule} that uses expert knowledge of existing features to pinpoint likely cases of collusion, ``$[\texttt{feedback amt} < a] \land [(\texttt{seller\_age} < b \lor \texttt{buyer\_age} < c)] \Rightarrow \texttt{collusion}$'' and ``$[(\texttt{gmv} - \texttt{price} > d] \land [\texttt{billing\_zip} \neq \texttt{delivery\_zip}] \Rightarrow \texttt{collusion}$'', where $a$, $b$, $c$, $d$ are hyperparameters. The business rules and hyperparameters are based on domain expertise and have been verified in the graph.
\end{itemize}
% \bo{PCS implementation details here etc, results analysis added to later part about pcs}

\textit{PCS implementation.} To improve inductive generalization and mitigate uncertainty, we implement weight noise ensembling into \namenospace, following the PCS framework. During the statistical learning component, we add random perturbations sampled from a zero-mean Gaussian distribution with a noise scale of 0.1 to the weights of each GNN model. In the reasoning component, we perform ten forward passes through the GCN for each batch during the E-step of the variational EM algorithm, accumulating the loss and rule scores from each sample and computing their average to obtain a more robust estimate of the variational posterior. As demonstrated by our empirical results, incorporating weight noise ensembling within the PCS framework improves \namenospace's generalization performance and calibration. We measure calibration with the Expected Calibration Error (ECE), which quantifies the difference between the model's predicted probabilities and the observed frequencies of correct predictions. A lower ECE indicates better calibration and reliability.
% \bo{define ECE, x and y axis of fig 6. expectation}

\begin{figure}
    \centering
    \hspace{-4mm}
    \includegraphics[width=0.5\textwidth]{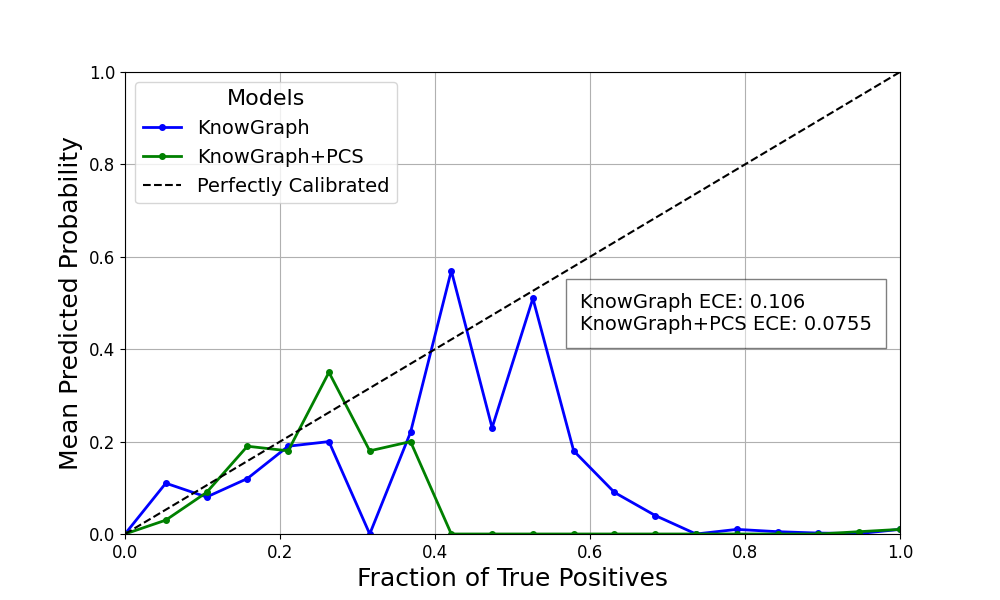}
    \caption{Model calibration for the \name framework with and without weight noise ensembling for the data-driven models. \name shows a more volatile calibration curve, deviating significantly from the dashed line representing perfect calibration. In contrast, \namenospace+PCS maintains a closer alignment to the perfectly calibrated line, suggesting more reliable probability estimates. The ECE values for \name (0.106) and \name+PCS (0.0755) quantify these observations, with the latter indicating a more accurately calibrated model.}
    \label{fig:pcs_base}
\end{figure}

\paragraph{Main Results} Combining model predictions with the main model in the reasoning framework improves performance in both the transductive and inductive settings. Notably, in Tab.~\ref{tab:overall} we observe a 0.086 AUC gain and 0.032 AP gain in the transductive setting over the individual main model DGI \cite{veličković2018deep}. This is also comparable to the highest performing baseline for the transductive setting, a GCN \cite{kipf2017semisupervised} trained with supervised learning. There is a more significant gain in the inductive setting, where we observe a 0.111 AUC and 0.120 AP improvement. Furthermore, adding weight noise ensembling from PCS further improves inductive generalization performance to 0.649 AUC and 0.167 AP, a 0.151 and 0.166 gain over the baseline, and the highest performance on this dataset. However, using PCS slightly lowers transductive AUC and AP but still outperforms baselines while improving TP at lower FP rates.

In addition, we find in Fig.~\ref{fig:auc_chart} that this performance improvement is consistent over various overlapping days, including the fully inductive setting where the model is evaluated on a new graph at a 20-day time shift. Performance appears to be \textit{independent} from the degree the graph shifts, indicating the learned features and rules are robust to changes in the graph in future days. Since it is prohibitively expensive to label new nodes, this is the first time a fully automated pipeline can classify new transactions on this dataset with the potential to assist human experts. However, due to the low performance of the main model, there is still a large gap between transductive and inductive performance, which we leave for future work.

\paragraph{Ablation Studies}

The reasoning component of \method~ learns a weight for each rule corresponding to their significance on the final classification decision of the framework. We list the associated weight for each rule in Tab.~\ref{tab:model-weights}. Rules with associated weights with higher magnitudes have a larger effect on inference. 

%{todo - add details about calibration and bins and ece, plot both lines in one curve}
Besides inductive generalization ability, it is also important for models to be well-calibrated; the probability associated with the predicted class label should
reflect its ground truth correctness likelihood. We conduct a calibration analysis shown in Fig. \ref{fig:pcs_base}. In this diagram, the x-axis represents the model's predicted probability (confidence), divided into twenty bins. The y-axis represents the actual fraction of positive instances within each bin. We find in Fig. \ref{fig:pcs_base} that the base \name exhibits higher variability in its predicted probabilities with a noticeable peak around the 0.6 mark on the fraction of true positives, indicating a point of overconfidence in its predictions. In contrast, \name with PCS remains closer to the perfectly calibrated line, suggesting that it is better calibrated and more consistent in its predictive probabilities across different thresholds. This is also supported by the ECE values, where adding PCS reduces ECE from 0.106 to 0.0755.

% \begin{itemize}
%     \item $[\text{subclass} = \text{collusion}]\Rightarrow [\text{main} = \text{collusion}]$
%     \item $[\text{collusion}_i = \text{true}] \Rightarrow [(\text{collusion}_{i-1} = \text{false} \land \text{collusion}_{i+1} = \text{false} \land \dots \land \text{collusion}_n = \text{false})]$
%     \item $[\text{main} = \text{collusion}] \Rightarrow [\text{main2} = \text{collusion}] \land [\text{main2} = \text{collusion}] \Rightarrow [\text{main} = \text{collusion}]$
%     \item $[\text{feedback} < a] \land [(\text{seller\_age} < b \lor \text{buyer\_age} < c)] \Rightarrow \text{collusion}$
%     \item $[(\text{gmv} - \text{price} > d] \land [\text{billing\_zip} \neq \text{delivery\_zip}] \Rightarrow \text{collusion}$
    
% \end{itemize}

\section{Discussion and Conclusion}

We propose \name, the first framework that integrates knowledge reasoning with GNNs for enhanced graph-based anomaly detection. \name combines multiple GNN models operating on different graph structures with a probabilistic logical reasoning component. We employ the PCS framework and introduce weight noise ensembling to mitigate uncertainty and improve the inductive generalization ability of our model. \name is flexible to various amounts and designs of rules, highly scalable to large graphs, and demonstrates robust performance in challenging scenarios with class imbalance and heterogeneous information. Experiments on two large-scale real-world datasets, an eBay dataset for collusion detection and the LANL network event dataset for intrusion detection, demonstrate \namenospace's superior performance over state-of-the-art GNN baselines in both transductive and inductive settings. These real-world graph datasets present significant challenges, such as extreme class imbalance, heterogeneous information, and the need for inductive generalization to new graphs. \namenospace's ability to effectively address these challenges highlights the potential of integrating domain knowledge into data-driven models for high-stakes, graph-based security applications.

\namenospace's modular architecture paves the way for more accurate and interpretable graph learning systems, bridging the gap between symbolic and neural approaches to graph learning. The reasoning approach introduced in this paper opens up new opportunities for leveraging domain knowledge in various real-world applications. Future research directions include exploring more advanced GNN architectures and investigating alternative inference techniques. Additionally, developing more efficient and scalable reasoning components could further enhance the applicability of \name to even larger real-world graphs.

\section{Acknowledgements}
This work is partially supported by the National Science
Foundation under grant No. 2046726,
NSF AI Institute ACTION No. IIS-2229876, DARPA GARD, the National Aeronautics and Space Administration (NASA) under grant No. 80NSSC20M0229, the Alfred P. Sloan Fellowship, the Meta research award, and the eBay research award.

% In this paper, we propose \method, a novel framework that integrates domain knowledge into Graph Neural Networks (GNNs) via sensing-reasoning. By combining a learning component with multiple knowledge models and a reasoning component that performs probabilistic logic inference using Markov Logic Networks (MLNs), \method~ enables flexible incorporation of symbolic domain knowledge into GNNs. Experiments on real-world applications demonstrate \method's superior performance over state-of-the-art GNNs in both transductive and inductive settings.

% \bo{maybe emphasize the real world graph data detection challenges here}

% \method's modular architecture paves the way for more accurate and interpretable graph learning systems, bridging the gap between symbolic and neural approaches to graph learning. Future research directions include exploring more advanced architectures, investigating alternative inference techniques, and extending \method~ to handle other graph learning tasks. The sensing-reasoning approach introduced in this paper opens up new opportunities for leveraging domain knowledge in various real-world applications.
% \clearpage
\bibliographystyle{ACM-Reference-Format}
\bibliography{main}

%%
%% If your work has an appendix, this is the place to put it.
\appendix

\end{document}

%% file: tables/lanl_descriptions.tex
\begin{table*}[t]
    \centering
    \caption{An overview of the learning models and rules designed for lateral movement detection on LANL \cite{kent2016cyber}, which we formulate as a link prediction problem to identify malicious edges. We design three models and three knowledge rules to ensure resilient inference of the base model. The rules are designed from domain knowledge and verified in the dataset.}
    \begin{tabular}{p{0.15\textwidth}p{0.35\textwidth}p{0.35\textwidth}}
    \toprule
    \textbf{Model} & \textbf{Description} & \textbf{Rule} \\
    \midrule
    Main & Main model to predict if an edge is malicious & Used in all rules as the overall base model \\
    \midrule
    Auth & Binary edge classifier to determine if the authentication type of an edge is NTLM & Authentication rule: ``$[\texttt{main} = \texttt{Mal}] \Rightarrow [\texttt{auth} = \texttt{NTLM}]$'' \\
    \midrule
    2-hop & Model to predict if a edge is malicious using a 2-hop enclosing graph  & Subgraph rules: ``$[\texttt{EncG} = \texttt{Mal}] \Rightarrow [\texttt{main} = \texttt{Mal}]$'', ``$[\texttt{EncG} = \texttt{Benign}] \Rightarrow [\texttt{main} = \texttt{Benign}]$'' \\
    \bottomrule
    \end{tabular}
    \label{tab:components}
\end{table*}

%% file: tables/lanl-overall.tex
% \begin{table}
%   \centering
%   \caption{Detection AUC/Average Precision of the baselines and our sensing-reasoning framework. We observe better performance on both transductive and inductive settings. }
%   \label{tab:lanl-overall}
%     \begin{tabular}{l|cccc}
%     \toprule
%     Method & Transductive & Inductive \\
%     \midrule
%     Euler~\cite{king2023euler}  & 0.9946 / 0.0433 & 0.8973 / 0.0193 \\
%     % \texttt{auth}  & 0.9440 / - & 0.7950 / - \\
%     EncG~\cite{zhang2018link}  & 0.9995 / 0.3249 & 0.8269 / 0.0338 \\
%     % \texttt{main}+\texttt{auth} & 0.9947 / 0.0403 & 0.9003 / 0.0233 \\
%     % \texttt{main}+\texttt{2hop} & 0.9997 / 0.5320 & 0.9006 / 0.0583 \\
%     Reasoning (Ours) & \bf 0.9999 / 0.8886 & \bf 0.9112 / 0.0852 \\
%     \bottomrule
%     \end{tabular}
% \end{table}

\begin{table}
  \centering
  \caption{Detection Performance of the baselines and our \name framework. We observe better performance for \name in both transductive and inductive settings. Transductive settings involve learning and testing on the same graph, leveraging its entire structure, whereas inductive settings require the model to generalize to new graphs or unseen parts of a graph based on learned patterns.}
  \label{tab:lanl-overall}
  \setlength{\tabcolsep}{3pt}
    \begin{tabular}{l|ccccc}
    \toprule
    \multicolumn{5}{c}{Transductive} \\
    \midrule
    Method & AUC & AP & TP@0.5FP & TP@1FP & TP@2FP\\
    \midrule
    Euler~\cite{king2023euler}  & 0.9946 & 0.0433 & 0.7777 & 0.9444 & \bf 1.0 \\
    EncG~\cite{zhang2018link}  & 0.9995 & 0.3249 & \bf 1.0 & \bf 1.0 & \bf 1.0 \\
    \name &  \bf 0.9999 & \bf 0.8886 & \bf 1.0 & \bf 1.0 & \bf 1.0 \\
    \midrule
    \multicolumn{5}{c}{Inductive} \\
    \midrule
    Method & AUC & AP & TP@0.5FP & TP@1FP & TP@2FP\\
    \midrule
    Euler~\cite{king2023euler}  &  0.8973 & 0.0193 & 0.0 & 0.0 & 0.2534 \\
    EncG~\cite{zhang2018link}  &  0.8269 & 0.0338 & 0.1648 & 0.2755 & 0.3950\\
    \name &  \bf 0.9112 & \bf 0.0852 & \bf 0.3554 & \bf 0.4643 & \bf 0.5910\\
    \bottomrule
    \end{tabular}
\end{table}

%% file: tables/lanl-abl.tex
\begin{table}
  \centering
  \caption{Detection performance (AUC) of the models and \name framework with partial knowledge. We observe that \name with the final reasoning framework indeed achieves the best detection performance.}
  \label{tab:lanl-abl}
    \begin{tabular}{l|cccc}
    \toprule
    Model & Transductive & Inductive \\
    \midrule
    \texttt{main}  & 0.9946 & 0.8973  \\
    \texttt{auth}  & 0.9440 & 0.7950 \\
    \texttt{EncG}  & 0.9995 & 0.8269 \\
    \texttt{main}+\texttt{auth} & 0.9947 & 0.9003 \\
    \texttt{main}+\texttt{EncG} & 0.9997 & 0.9006 \\
    \texttt{main}+\texttt{auth}+\texttt{EncG} & \bf 0.9999 & \bf 0.9112 \\
    \bottomrule
    \end{tabular}
\end{table}

%% file: tables/ebay_descriptions.tex
\begin{table*}[t]
    \centering
     \caption{An overview of the models and knowledge rules designed for collusion detection on the real-world eBay dataset, a heterogeneous graph of marketplace transactions. We design a total of eight data-driven learning models and six knowledge rules to enhance the inference of the main task model. These rules are designed by experts with domain knowledge.}
    \begin{tabular}{p{0.15\textwidth}p{0.35\textwidth}p{0.35\textwidth}}
    \toprule
    \textbf{Model} & \textbf{Description} & \textbf{Knowledge Rule} \\
    \midrule
    Main & Main model to predict if a user, transaction, or item node is an instance of collusion & Used in all rules as the overall base model \\
    \midrule
    Secondary & Same objective as the main model, but trained on a larger graph snapshot with balanced sampling & Used in the Ensemble Rules: ``$[\texttt{main} = \texttt{collusion}]  \Rightarrow [\texttt{secondary} = \texttt{collusion}]$'' and ``$[\texttt{secondary} = \texttt{collusion}] \Rightarrow [\texttt{main} = \texttt{collusion}]$.'' \\
    \midrule
    Account Takeover & Model to predict collusion subclass where one of the colluding parties fraudulently takes control of another account & Used in Hierarchy Rule: ``$[\texttt{ATO} = \texttt{collusion}]\Rightarrow [\texttt{main} = \texttt{collusion}]$.''  \\
    \midrule
    Sign-In & Model to predict collusion subclass accompanied by some problem indicator related to sign-in & Used in Hierarchy Rule: ``$[\texttt{sign-in} = \texttt{collusion}]\Rightarrow [\texttt{main} = \texttt{collusion}]$.'' \\
    \midrule
    One Day Reg & Model to predict collusion subclass where buyers register and purchase an item on the same day & Used in Hierarchy Rule: ``$[\texttt{one-day-reg} = \texttt{collusion}]\Rightarrow [\texttt{main} = \texttt{collusion}]$.'' \\
    \midrule
    US eBay Decline & Model to predict collusion subclass where the transaction is declined by eBay but overridden by the colluding party & Used in Hierarchy Rule, ``$[\texttt{ebay-decline} = \texttt{collusion}]\Rightarrow [\texttt{main} = \texttt{collusion}]$.'' \\
    \midrule
    Account Age & Domain knowledge using labeled attributes on the age and amount of user feedback of an account & Used in Business Rule, ``$[\texttt{feedback amt} < a] \land [(\texttt{seller\_age} < b \lor \texttt{buyer\_age} < c)] \Rightarrow \texttt{collusion}$'' \\
    \midrule
    Business & Domain knowledge using labeled attributes of the zip code, price, and gross value of a transaction & Used in Business Rule, ``$[(\texttt{gmv} - \texttt{price} > d] \land [\texttt{billing\_zip} \neq \texttt{delivery\_zip}] \Rightarrow \texttt{collusion}$'' \\
    \bottomrule
    \end{tabular}
    \label{tab:components}
\end{table*}

%% file: tables/ebay_overall.tex
% \begin{table}
%   \centering
%   \caption{Overall performance of baselines and \name. We observe significantly higher performance for \name in both transductive and inductive settings. We report AUC and average precision with $k=0.5$.}
%     \begin{tabular}{l|cccc}
%     \toprule
%     Sensor & Transductive & Inductive \\
%     \midrule
%     GCN {\scriptsize~\cite{kipf2017semisupervised}} & 0.9531 & 0.501 \\
%     DGI {\scriptsize~\cite{veličković2018deep}} & 0.892 & 0.498 \\
%     \rowcolor{gray!20}\textbf{\name} & \textbf{0.978} & 0.609 \\
%     \rowcolor{gray!20}\textbf{+ PCS} & 0.952 & \textbf{0.649} \\
%     \bottomrule
%     \end{tabular}
%   \label{tab:overall}
% \end{table}

\begin{table}
  \centering
  \caption{Overall performance of baselines and \namenospace. \name outperformes the baseline DGI \cite{veličković2018deep}  in both the transductive and inductive settings. The performance is further improved with PCS framework. We report AUC and average precision with $k=0.5$, where $k$ is the proportion of the total dataset considered when evaluating average precision. We bold results with an improvement larger than 1\%.}
  \label{tab:overall}
  \setlength{\tabcolsep}{2.5pt}
    \begin{tabular}{l|ccccc}
    \toprule
    \multicolumn{5}{c}{Transductive} \\
    \midrule
    Method & AUC & AP & TP@0.5FP & TP@1FP & TP@2FP\\
    \midrule
    GCN {\cite{kipf2017semisupervised}} & 0.953 & 0.723 & 0.719 & 0.742 & 0.790 \\
    DGI {\cite{veličković2018deep}}  & 0.892 & 0.587 & 0.717 & 0.738 & 0.788 \\
    \rowcolor{gray!20}\textbf{\name} & 0.978 & \bf 0.755 & 0.714 & 0.745 & 0.788 \\
    \rowcolor{gray!20}\textbf{\namenospace+PCS} & 0.968 & 0.731 & \bf 0.731 & \bf 0.757 & \bf 0.814 \\
    \midrule
    \multicolumn{5}{c}{Inductive} \\
    \midrule
    Method & AUC & AP & TP@0.5FP & TP@1FP & TP@2FP\\
    \midrule
    GCN {\cite{kipf2017semisupervised}} & 0.501 & 0.0001 & 0.0 & 0.0096 & 0.0096 \\
    DGI {\cite{veličković2018deep}}  & 0.498 & 0.0001 & 0.0 & 0.0096 & 0.0096 \\
    \rowcolor{gray!20}\textbf{\namenospace} & 0.609 & 0.121 & 0.0048 & 0.0096 & 0.0144 \\
    \rowcolor{gray!20}\textbf{\namenospace+PCS} & \bf 0.649 & \bf 0.167 & \bf 0.0287 & \bf 0.0383 & \bf 0.0431 \\
    \bottomrule
    \end{tabular}
\end{table}

%Domain Knowledge & 0.531 & 0.0578 & 0.0666 & 0.0713 & 0.0807 \\
%Domain Knowledge & 0.5335 & 0.0672 & \bf 0.0717 & \bf 0.0763 & \bf 0.0856 \\

%% file: tables/ebay_rule_weights.tex
\begin{table*}[t]
\centering
\caption{Final rule weights of \method~applied to collusion detection on the real-world eBay dataset. We observe similar weights in similar rules, such as the hierarchy rule. Rules with higher magnitude have a larger overall effect on inference.}
\label{tab:model-weights}
\begin{tabular}{@{}lrrrrrrrr@{}}
\toprule
& ATO & Sign-in & One Day Reg & US Ebay Decline & Exclusion & Ensemble & Account & Price/Shipping \\
\midrule
Weight & -1.2227e-02 & -1.2221e-02 & -1.2221e-02 & -1.2221e-02 & -2.2177e-10 & -1.5172e-02 & 1.6767e-03 & 1.5844e-03 \\
\bottomrule
\end{tabular}
\end{table*}

% Weight & -1.2227e-02 & -1.2221e-02 & -1.2221e-02 & -1.2221e-02 & -2.2177e-10 & -1.5172e-02 & 1.6767e-03 & 1.5844e-03 \\

%% file: tables/ebay_model.tex
\begin{table*}[!t]
\centering
\caption{Knowledge model performance under the transductive and inductive settings for the main task (first two rows) and their own semantic tasks (third row). We report average precision with $k = 0.5$. 
The knowledge models perform well on their own semantic tasks in general and perform poorly when directly mapped to the main task. 
% Most models also do not transfer to either the overall transductive or inductive setting, except the Account Age and Business models based on labeled features, which individually can classify several more identifiable collusion cases.
}
\label{tab:model-performance}
\begin{tabular}{@{}l|c|ccccc|cc@{}}
\toprule
Methods & Main & Secondary & ATO & Sign-in & One Day Reg & US Ebay Decline & Account Age & Business \\
\midrule
Transductive & 0.587 & 0.5805 & 0.0002 & 0.0011 & 0.0047 & 0.0014 & 0.0434 & 0.0378 \\
Inductive & 0.0001 & 0.0001 & 0.0001 & 0.0001 & 0.0001 &
0.0001 & 0.0526 & 0.0692 \\
\midrule
Knowledge Task & 0.587 & 0.5805 & 0.681 & 0.521 & 0.353 & 0.251 & 1.0 & 1.0 \\
\bottomrule
\end{tabular}
\end{table*}

%Knowledge Task & 0.5805 & 0.681 & 0.521 & 0.353 & 0.251 & 1.0 & 1.0 \\